\documentclass[aps,prd,twocolumn,showpacs,superscriptaddress,nofootinbib,nobibnotes,longbibliography]{revtex4-1}  \usepackage[latin1]{inputenc}
\usepackage{mathtools}
\usepackage{amsmath,amssymb}
\usepackage{amsfonts}
\usepackage{amsthm}
\usepackage{latexsym}
\usepackage{graphicx}
\usepackage{epstopdf}
\usepackage[footnotesize]{caption}
\usepackage{subfig}
\usepackage[colorlinks]{hyperref}

\newcommand{\note}[1]{\text{\tiny{#1}}}
\newcommand{\ee}{\mathrm{e}}
\newcommand{\ii}{\mathrm{i}}
\newcommand{\dd}{\mathrm{d}}
\newcommand*\DAlembert{\mathop{}\!\mathbin\Box}
\renewcommand{\rho}{\varrho}
\renewcommand{\epsilon}{\varepsilon}

\theoremstyle{plain}

\begin{document}

\title{Constructing Neutron Stars with a Gravitational Higgs Mechanism}
\author{Nicola Franchini}
\affiliation{School of Mathematical Sciences, University of Nottingham, University Park, Nottingham, NG7 2RD, UK}
\author{Andrew Coates}
\affiliation{School of Mathematical Sciences, University of Nottingham, University Park, Nottingham, NG7 2RD, UK}
\author{Thomas P. Sotiriou}
\affiliation{School of Mathematical Sciences, University of Nottingham, University Park, Nottingham, NG7 2RD, UK}
\affiliation{School of Physics and Astronomy, University of Nottingham, University Park, Nottingham, NG7 2RD, UK}

\begin{abstract}
In scalar-tensor theories, spontaneous scalarization is a phase transition that can occur in ultra-dense environments such as neutron stars. The scalar field develops a non-trivial configuration once the stars exceeds a compactness threshold. We recently pointed out that, if the scalar exhibits some additional coupling to matter, it could give rise to significantly different microphysics in these environments. In this work we study, at the non-perturbative level, a toy model in which the photon is given a large mass when spontaneous scalarization occurs. Our results  demonstrate clearly the effectiveness of spontaneous scalarization as a Higgs-like mechanism in neutron stars.
\end{abstract}

\maketitle

\section{Introduction}

Scalar-tensor theories \cite{Faraoni:2004pi,Fujii:2003pa,Damour:1992we} are probably the most studied alternative theories of gravity. They can be described by the action
\begin{align}\label{eq:actionJF}
S_J= &\frac{1}{16\pi}\int \dd^4 x\sqrt{-\tilde{g}}\left(\Phi \tilde{R}-\frac{\omega(\Phi)}{\Phi}\tilde{\nabla}^\mu\Phi\partial_\mu\Phi\right)\nonumber\\
&+ S_m[\Psi^A,\tilde{g}_{\mu\nu}] ,
\end{align}
where $\tilde{R}$ is the Ricci scalar of $\tilde{g}_{\mu\nu}$ and $S_m$ is the matter action for generic matter fields $\Psi^A$. In this representation, known as the Jordan frame, the scalar field is non-minimally coupled to the gravitational field but the matter fields couple minimally to the metric only. An alternative representation is that of the Einstein frame, where the action reads
\begin{align}\label{eq:actionEF}
S_E= \frac{1}{16\pi G_*}\int\dd^4x\sqrt{-g}(R-2\nabla^\mu\phi\partial_\mu\phi) \nonumber\\
 +S_m[\Psi^A,B^2(\phi)g_{\mu\nu}].
\end{align}
The Einstein frame metric $g_{\mu\nu}$ and the scalar field $\phi$ are related to their Jordan frame counterparts via the relations
\begin{equation}\label{eq:conftransf}
\Phi=(G_*B^2(\phi))^{-1}, \qquad \tilde{g}_{\mu\nu}=B^2(\phi)g_{\mu\nu},
\end{equation}
and $\omega(\Phi)$ is mapped to $B(\phi)$ through the relation
\begin{equation}\label{eq:omegaB}
\alpha(\phi)\equiv\frac{\dd}{\dd\phi}\log B(\phi)=-\sqrt{\frac{1}{2\omega(\Phi)+3}}.
\end{equation}
In the Einstein frame $\phi$ couples minimally to $g_{\mu\nu}$ and it has a canonical kinetic term (up to a constant field rescaling), so in the absence of matter fields the theory reduces to general relativity. However, now $\phi$ is coupled to matter fields when they are present. One can also include a potential for either $\Phi$ or $\phi$, but this will not be necessary for our purposes.

Clearly, one can perform field redefinitions and change the representation of the theory, without changing the underlying physics. However, associating a given equation (or action) with a physical system requires one to relate the variables of a given representation with characteristics of the system \cite{Sotiriou:2007zu}. As an example, if one is given the equation of a harmonic oscillator written in terms of a variable $x$, one needs to know if $x$ is the displacement of the oscillator in order to ascribe a physical meaning to the solutions. One can clearly redefine $x$ and change the form of the differential equation, but the physical interpretation of the results remains unambiguous. In scalar-tensor theories, in order to ascribe meaning to the fields and distinguish between representations, one resorts to the weak equivalence principle or universality of free fall. The latter asserts that test bodies of different composition follow the same trajectories when in free fall. This is satisfied if test bodies follow geodesics of a metric. In the context of scalar tensor theories, this metric is $\tilde{g}_{\mu\nu}$ because the matter fields $\Psi^A$ couple minimally to it. Whereas in the Einstein frame, the matter fields couple to both $g_{\mu\nu}$ and $\phi$ and hence they will not follow geodesics of $g_{\mu\nu}$ because they will experience a (fifth) force mediated by $\phi$. We stress that these are two variable-dependent interpretations that are fully compatible and describe precisely the same physics \cite{Sotiriou:2007zu}.

The above reasoning singles out the Jordan frame and suggests that its existence is necessary in order to satisfy the weak equivalence principle.\footnote{An obvious caveat to this line of thinking is that geodesic motion is not a necessary but just a sufficient condition for universality of free fall. However, we are restricting ourselves to metric theories in general and scalar-~tensor theories in particular.} However, it relies heavily on the implicit assumption that $\phi$ has a nonzero gradient in the environment one is testing the equivalence principle. If $\phi$ is in a trivial configuration with zero gradient it will not exert any fifth force on matter. This can also be seen by the fact that, when $\phi$ is constant, $g_{\mu\nu}$ and $\tilde{g}_{\mu\nu}$ differ just by a constant rescaling, so their geodesics coincide.

This caveat is not relevant for most scalar-tensor theories, as they do not actually admit $\phi=$constant solutions in the presence of matter. This can be seen directly from the field equations in the Einstein frame
\begin{align}\label{eq:einstein}
R_{\mu\nu}-\frac{1}{2}g_{\mu\nu}R&=8\pi G_* \left(T_{\mu\nu}+T^{(\phi)}_{\mu\nu}\right),\\
\label{eq:scalarfield}
  \DAlembert\phi&=-4\pi G_* T\alpha(\phi),
\end{align}
with $\alpha(\phi)$ defined in Eq.~\eqref{eq:omegaB}, $T$ is the trace of the Einstein frame stress-energy tensor
\begin{equation}\label{eq:stressenergytensor}
T_{\mu\nu}=-\frac{2}{\sqrt{-g}}\frac{\delta S_m}{\delta g^{\mu\nu}}\,,
\end{equation}
and
\begin{equation}
T^{(\phi)}_{\mu\nu}=\frac{1}{4\pi G_*}\left(\partial_\mu\phi\partial_\nu\phi-\frac{1}{2}g_{\mu\nu}\left(\partial_\alpha\phi\right)\left(\partial^\alpha\phi\right)\right).
\end{equation}
Only the class of theories for which $\alpha(\phi_0)=0$ for some constant $\phi_0$ will admit constant $\phi$ solutions when $T\neq 0$. These solution will actually be solutions of general relativity \cite{Sotiriou:2015lxa}.

Remarkably, the same class of theories can exhibit what is know as spontaneous scalarization \cite{Damour:1993hw}. Though the $\phi=\phi_0$ solutions exist, they do not have to be unique. In matter configurations of lower compactness, which should at least include the Sun to satisfy solar system constraints, this solution is indeed energetically favoured and the theory completely resembles general relativity. However, once the compactness of the matter configuration exceeds a certain threshold, whose value depends on the value of $\beta$, a non-trivial configuration for $\phi$ becomes favourable. This  leads to drastic differences with respect to GR \cite{Damour:1993hw}.
In particular, one can expand $\alpha(\phi)$ around the constant solution $\phi_0$ as
\begin{equation}\label{eq:expalpha}
\alpha(\phi)=\alpha_0+\beta(\phi-\phi_0)+\dots
\end{equation}
Setting $\alpha_0=0$ would lead to exact agreement with general relativity in Solar System experiments \cite{EspositoFarese:2004cc}.The term on the right hand side of Eq.~\eqref{eq:scalarfield} gives rise to an effective mass for linear perturbations of the scalar field. The square of the effective mass is $-4\pi G_*\beta T$ and it has the wrong sign for negative values of $\beta$ and \(T<0\). Indeed, at perturbative level, spontaneous scalarization manifests itself as a tachyonic instability of the $\phi=\phi_0$ solution. Numerical simulations show that $\beta\lesssim-4.35$ in order to  generate a nontrivial scalar profile for neutron stars \cite{Harada:1997mr,Chiba:1997ms}.

Spontaneous scalarization is usually thought of as a phase transition that induces a scalar charge for compact stars. However, recalling that compactness determines curvature, there is an equivalent but alternative perspective which we find enlightening for our discussion. The theories that exhibit spontaneous scalarization have an extra dynamical scalar degree of freedom. In the large curvature regime this scalar leads to new phenomenology. In the small curvature regime, it develops a steep effective potential that strongly stabilises it to its minimum and makes it hard to excite. As a result, this scalar is ``screened''  and hard to detect in low curvature environments.

With this perspective in mind, it is not hard to see how a theory that exhibits such behaviour does not need to have a Jordan frame to satisfy equivalence principle tests. Consider adding to action \eqref{eq:actionJF} an extra interaction piece $S_{\rm int}[\Psi^A,\tilde{g}_{\mu\nu},\Phi]$ that directly couples $\Phi$ to the matter fields $\Psi^A$. Recall that $\phi$=constant implies $\Phi=$constant.  If $S_{\rm int}$ is designed so that
\begin{enumerate}
\item[(i)] it vanishes for $\Phi=$constant and so do its contributions to the field equations,
\item[(ii)] it does not alter the scalarization picture described above,
\end{enumerate}
then the corresponding theory would not actually have a Jordan frame but it would also not exhibit a violation of the weak equivalence principle in the Solar System. It would continue to be indistinguishable from general relativity in the weak field. Remarkably though, in the strong field regime and in the interior and the vicinity of compact stars where the scalar field will be non-trivial, one will not only alter the metric, but there will also be modification to the standard model of particle physics thanks to the extra interactions between the matter fields and the scalar.

Models that satisfy the criteria (i) and (ii) above and yet lead to new physics in and around neutron stars do exist, as clearly demonstrated in Ref.~\cite{Coates:2016ktu}.  In particular, it was shown there that, by suitably coupling a (complex) scalar that undergoes scalarization in high curvature to the electromagnetic field, one can generate a mass for the photon (for more details about spontaneous scalarization with a complex scalar field see Ref.~\cite{Horbatsch:2015bua}). The mechanism resembles the Higgs mechanism and the model of Ref.~\cite{Coates:2016ktu} was intentionally simple and provocative. In principle one can use exactly the same mechanism to modify other aspects of the standard model in the high curvature regime.

The take-away message from all of the above is the following: there might be extra degrees of freedom in gravity which are actually screened in the weak gravity regime that we usually probe. If this screening is efficient enough it can actually suppress couplings between these new degrees of freedom and matter fields. This would allow the matter physics in the interior of compact stars to be different than the physics in our laboratories and accelerators. This possibility is usually ignored. It could have important consequences for our perception of what constitutes a realistic equation of state for compact stars and it might lead to exciting new phenomena.

The analysis of Ref.~\cite{Coates:2016ktu} was perturbative. As mentioned earlier, at perturbative level scalarization manifests itself as a tachyonic instability and one can study this instability in order to demonstrate that the model under consideration will exhibit spontaneous scalarization and assess whether it will satisfy criterion (ii) above. However, the perturbative treatment  does not allow one to probe the end-state of the tachyonic instability and, hence, it does not provide any description of the scalarized configuration. Therefore, one cannot use it to assess how efficient scalarization is in changing the properties of matter in and around the star. This is clearly very important. An objection one might have to adding a coupling between matter and $\phi$ is that, even if the coupling terms vanish when $\phi=\phi_0$, there will still be new interaction vertices.  This would lead to new phenomenology that should show up in accelerators, unless the coupling is very weak. The latter can always be arranged but then one needs to explicitly show that scalarization can enhance this interaction enough to actually change the physics in the interior of the star.

The purpose of this paper is to revisit the model of Ref.~\cite{Coates:2016ktu} and perform a non-perturbative analysis. We will construct static, spherically symmetric neutron star solutions and we will show that spontaneous scalarization does indeed go through as expected and it endows the photon with a $\phi$-dependent mass. As a result, the mass of the photon has a radial profile. Remarkably, the mass can become very large even for very small values of the charge that controls the interaction between $\phi$ and the electromagnetic field.

In the next section we review the model and in Sec. \ref{sec:setup} we will lay out the setup of the problem. In Sec. \ref{sec:emfield} we will explore the electromagnetic field configuration and in Sec. \ref{sec:mass} we will present out numerical solutions for scalarized neutron stars and the corresponding photon mass profile. Sec. \ref{sec:disc} contains a discussion of our results and future prospects.

\section{The model}

We will focus on the model introduced in Ref.~\cite{Coates:2016ktu}
\begin{equation}\label{eq:actionEFEM}
\begin{multlined}
S_E=\int\dd^4x\sqrt{-g}\left(\frac{R}{4}-\frac{1}{2}g^{\mu\nu}\overline{D_\mu \phi}D_\nu \phi\right)+ \\
 -\frac{1}{4}\int\dd^4x\sqrt{-g}F_{\mu\nu}F^{\mu\nu}+ \\
 +S_m[\Psi^A;B^2(\bar{\phi}\phi)g_{\mu\nu};A_\mu],
\end{multlined}
\end{equation}
where $F_{\mu\nu}=A_{\nu,\mu}-A_{\mu,\nu}$ is the electromagnetic tensor, $D_\mu \phi=\phi_{,\mu}-\ii e A_\mu\phi$ is the gauge covariant derivative of the scalar, and $e$ is the coupling charge (hereafter we set $4\pi G_*=c=\hbar=1$). The scalar field is complex and one can check that the action is invariant under the gauge transformation
\begin{equation}\label{eq:gauge}
\phi\rightarrow\phi\ee^{\ii e\lambda},\qquad A_\mu\rightarrow A_\mu+\lambda_{,\mu}.
\end{equation}
If this symmetry is spontaneously broken, the photon acquires  a non-zero mass. The action is given in the Einstein frame. Using the redefinition of Eq.~\eqref{eq:omegaB} with $B^2(\bar{\phi}\phi)$, one can write an action in a different conformal frame that will correspond to the Jordan frame when $e=0$. For any nonzero value of $e$, there will still be a coupling between $A_\mu$ and $\phi$ in this frame. For want of a better name and with an abuse of terminology,  in what follows we will refer to this frame as the Jordan frame. It is worth emphasising that all matter fields other than $A_\mu$ do couple minimally to the metric only in this frame. The mass of the photon in the Einstein and in the Jordan frame are
\begin{equation}\label{eq:photonmass}
m_{\gamma\note{,E}}^2(\bar{\phi}\phi)=e^2\bar{\phi}\phi, \qquad m_{\gamma\note{,J}}^2(\bar{\phi}\phi)=\ee^{-\beta\bar{\phi}\phi}e^2\bar{\phi}\phi.
\end{equation}

It is convenient to split the scalar field into its real and imaginary parts, i.e. $\phi=\phi_1+\ii\phi_2$. The variation of the action with respect to $\phi_1$, $\phi_2$, $A^\nu$ and $g^{\mu\nu}$ yields the following equations
\begin{gather}
\begin{aligned}
\left(\DAlembert-e^2A_\mu A^\mu\right)\phi_1 &+\left(2e A^\mu\partial_\mu+e\nabla_\mu A^\mu\right)\phi_2= \\
=&-T \frac{\dd}{\dd\phi}\log B(|\phi|^2)\phi_1  \label{eq:phi1}
\end{aligned}\\
\begin{aligned}
\left(\DAlembert-e^2A_\mu A^\mu\right)\phi_2 &-\left(2e A^\mu\partial_\mu+e\nabla_\mu A^\mu\right)\phi_1= \\
=&-T \frac{\dd}{\dd\phi}\log B(|\phi|^2)\phi_2 \label{eq:phi2}
\end{aligned}\\
\nabla^\mu F_{\mu\nu}=J_\nu+J_\nu^{(\phi)}+m^2_\gamma(|\phi|^2) A_\nu, \label{eq:EM}\\
\label{eq:einsEM} G_{\mu\nu}= 2\left(T_{\mu\nu}+T_{\mu\nu}^{(\phi)}+T_{\mu\nu}^{(A)}+T_{\mu\nu}^{(\phi A)}\right),
\end{gather}
where
\begin{gather}\label{eq:SEtensor}
\begin{aligned}
T_{\mu\nu}^{(\phi)}=\phi_{1,\mu}\phi_{1,\nu}&+\phi_{2,\mu}\phi_{2,\nu}+\\
&- \frac{1}{2}g_{\mu\nu}g^{\rho\sigma}\left(\phi_{1,\rho}\phi_{1,\sigma}+ \phi_{2,\rho}\phi_{2,\sigma}\right),
\end{aligned}\\
\begin{aligned}
T_{\mu\nu}^{(A)}=F_{\mu\rho}F_\nu^\rho & -\frac{1}{4}g_{\mu\nu}F_{\rho\sigma}F^{\rho\sigma}+ \\
& +m^2_\gamma\left(A_\mu A_\nu-\frac{1}{2}g_{\mu\nu}A_\rho A^\rho\right),
\end{aligned}\\
J^{(\phi)}_\mu=e\left(\phi_2\phi_{1,\mu}-\phi_1\phi_{2,\mu}\right), \\
T_{\mu\nu}^{(A\phi)}=J^{(\phi)}_\mu A_\nu+J^{(\phi)}_\nu A_\mu-g_{\mu\nu}J^\phi_\rho A^\rho.
\end{gather}

\section{Setup}
\label{sec:setup}

Our goal is to obtain solutions that describe static, spherically symmetric stars. We will use the following ansatz for the metric
\begin{equation}\label{eq:ansatz1}
  ds^2=-\ee^{\nu(r)}\dd t^2+\frac{\dd r^2}{1-2\mu(r)/r}+r^2\dd\Omega^2.
\end{equation}
We will model all matter other than $A_\mu$ as a perfect fluid. Its stress energy tensor in the Jordan frame reads
\begin{equation}
\tilde{T}^{\mu\nu}=(\epsilon+p)u^\mu u^\nu+p\tilde{g}^{\mu\nu},
\end{equation}
and it is related to the Einstein frame stress energy tensor by
\begin{equation}
T^{\mu\nu}=B^6(\phi) \tilde{T}^{\mu\nu}.
\end{equation}
Let us first look at the equations in the more standard scalarization scenario \cite{Damour:1993hw}. Starting from Eqs.~ \eqref{eq:einstein}-\eqref{eq:scalarfield} and selecting $B(\phi)=\ee^{\frac{1}{2}\beta\phi^2}$, which corresponds to  $\alpha(\phi)=\beta \phi$, one obtains the following system of equations
\begin{equation} \label{eq:sist1}
\begin{aligned}
  \mu'= & r^2 \ee^{2\beta\phi^2}\epsilon+\frac{1}{2}r(r-2\mu)\psi^2,\\
  \nu'= & \frac{2r^2 \ee^{2\beta\phi^2}p}{r-2\mu}+r\psi^2+\frac{2\mu}{r(r-2\mu)},\\
  p'= & -(\epsilon+p)\left(\frac{\nu'}{2}+\beta\phi\psi\right), \\
  \phi'= & \psi,\\
  \psi'= & \frac{r \ee^{2\beta\phi^2}}{r-2\mu}[\beta\phi(\epsilon-3p)+ r\psi(\epsilon-p)]+ \\
  &-\frac{2(r-\mu)}{r(r-2\mu)}\psi.
\end{aligned}
\end{equation}
In our model, we need to make an ansatz for the electromagnetic field as well.  In spherical symmetry we can make the choice
\begin{equation}\label{eq:ansatzA}
A_\mu=\left(A_0(r),A_1(r),0,0\right),
\end{equation}
without loss of generality.
With the above ansaetze we note that the $\mu=r$ equation of \eqref{eq:EM} is a constraint for $\phi_1$ and $\phi_2$
\begin{equation}\label{eq:constraintphi}
e A_1|\phi|^2=\phi_1\phi_2'-\phi_2\phi_1'.
\end{equation}
With a straightforward calculation, one can see that substituting this constraint into either of the Eqs. \eqref{eq:phi1}~-~\eqref{eq:phi2} yields the same equation. This can be thought of a consequence of gauge invariance. Making the gauge choice $\phi_2=0$ automatically sets $A_1=0$. It is worth noting that with this rotation of the scalar, its field current $J^{(\phi)}_\mu$ vanishes everywhere, and thus $T_{\mu\nu}^{(A\phi)}$ also vanishes.

With these gauge conditions it is possible to write a system of equations analogous to system \eqref{eq:sist1}
\begin{equation}\label{eq:sistEM}
\begin{gathered}
  \mu'  =\mu'_\note{DEF}+\frac{1}{2}\ee^{-\nu}r\left[ e^2rA_0^2\phi^2+(r-2\mu)f^2 \right], \\
  \nu'  =\nu'_\note{DEF}+\frac{r\ee^{-\nu}}{r-2\mu}\left[ e^2rA_0^2\phi^2-(r-2\mu)f^2\right], \\
  p'  =p'_\note{DEF}-2e^2A_0\phi^2 f\ee^{-\nu-2\beta\phi^2}, \\
  \phi'  =\psi, \\
  \psi' = \psi'_\note{DEF}-\frac{\ee^{-\nu}r}{r-2\mu}\left[e^2A_0^2\phi-(r-2\mu)f^2\psi\right], \\
  A_0'  =f, \\
\begin{multlined}
  f'  = f\frac{r^3\ee^{2\beta\phi^2}(p+\epsilon)+(r-2\mu)(r^2\psi^2-2)}{r(r-2\mu)}+ \\
  + e^2_\phi rA_0\phi^2\,\frac{\ee^{-\nu}rA_0f-1}{r-2\mu},
\end{multlined}
\end{gathered}
\end{equation}
where a quantity with the ``DEF" label is equal to the corresponding quantity of system \eqref{eq:sist1}. Given the system \eqref{eq:sistEM} and an equation of state (EOS) which links pressure and density, it is possible to find  solutions that describe neutron stars.

\section{Electromagnetic field configuration}
\label{sec:emfield}

The system \eqref{eq:sistEM} admits solutions in which $A_0=f=0$ and they are also solutions of system \eqref{eq:sist1}. That is, every solution to the standard scalarization scenario is a solution to the models studied here with vanishing electromagnetic field. Naively, this might seem obvious, as we have assumed that the perfect fluid is electrically neutral and does not source the electromagnetic field. However, upon closer inspection of Eq.~\eqref{eq:EM} one sees that the electromagnetic field is sourced by the scalar current  $J^{(\phi)}_\mu$ as well. Above we explained how our gauge choice allows us to set this current to zero. The question that remains is whether the solutions where $A_0=f=0$ are actually unique.  In Ref.~\cite{Coates:2016ktu} it was shown that this is indeed the case at perturbative level. Below we prove it without any approximation, assuming only regularity at the centre and asymptotic flatness.

Let us consider Eq.~\eqref{eq:EM}. In the gauge where $J^{(\phi)}_\nu$ vanishes, it becomes the Proca field equation with a $\phi$-dependent mass
\begin{equation}
\nabla^\mu F_{\mu\nu}=m^2_\gamma(|\phi|^2) A_\nu, \label{eq:EM0}\\
\end{equation}
Contracting with $A^\nu$ and integrating it over a volume $\mathcal{V}$ yields
\begin{equation}\label{eq:integralEM1}
\int_\mathcal{V}\dd^4 x\sqrt{-g}\left(A^\nu\nabla^\mu F_{\mu\nu}-m_\gamma^2A^\nu A_\nu\right)=0.
\end{equation}
We can integrate the first term by parts to obtain
\begin{equation}\label{eq:integralEM2}
\int_\mathcal{V}\dd^4 x\sqrt{-g}\left(\frac{F^{\mu\nu}F_{\mu\nu}}{2}+m_\gamma^2 A^\nu A_\nu\right)=\int_{\partial\mathcal{V}}\dd^3 \sigma\,n^\mu A^\nu F_{\mu\nu},
\end{equation}
where $\partial {\cal V}$ is the boundary of ${\cal V}$. Assuming regularity throughout  spacetime, we can choose this boundary to be a surface of constant radius $r$ and take the limit where $r\to \infty$. Let us use the following short-hand notation for the integrands, ${\cal I}_1\equiv F^{\mu\nu}F_{\mu\nu}/2+m_\gamma^2 A^\nu A_\nu$ and ${\cal I}_2\equiv n^\mu A^\nu F_{\mu\nu}$, where we have deliberately left the metric coefficients implicit. They are both covariant scalars so we can determine their properties by evaluating them using the ansaetze in Eqs.~\eqref{eq:ansatz1} and \eqref{eq:ansatzA}.
One has ${\cal I}_1=g^{tt}(g^{rr}(\partial_r A_0)^2+m_\gamma^2 A_0^2)$ and ${\cal I}_2= g^{tt}A_0\partial_r A_0$. In the limit $r\rightarrow\infty$, asymptotic flatness dictates that $A_0\propto1/r$ and $g^{tt}\to -1$. The volume element scales as $r^2$ and ${\cal I}_2$ scales as $r^{-3}$, thus the right hand side of Eq.~\eqref{eq:integralEM2} must vanish. This implies that ${\cal I}_1$ must vanish as well, as it is sign-definite when $g^{tt}$ and $g^{rr}$ do not change sign. Hence,  $A_0=0$. It  should be noted that our proof follows the no-hair proof for Proca fields in black hole backgrounds first presented in Ref.~\cite{Bekenstein:1971hc}.

We have shown that the only solutions of the system \eqref{eq:sistEM} that are regular throughout and have the right asymptotic fall-off have vanishing electromagnetic field and are also solutions of the system \eqref{eq:sist1}. This simplifies our task significantly: in order to determine the mass that the  photon will acquire upon scalarization, one just needs to re-derive the known solutions of the system \eqref{eq:sist1} and use them to calculate the mass through Eq.~\eqref{eq:photonmass}. We will do this in the next section.

\section{Scalarized stars and photon mass profile}
\label{sec:mass}

\begin{figure}[t]
  \centering
  \includegraphics[width=\columnwidth]{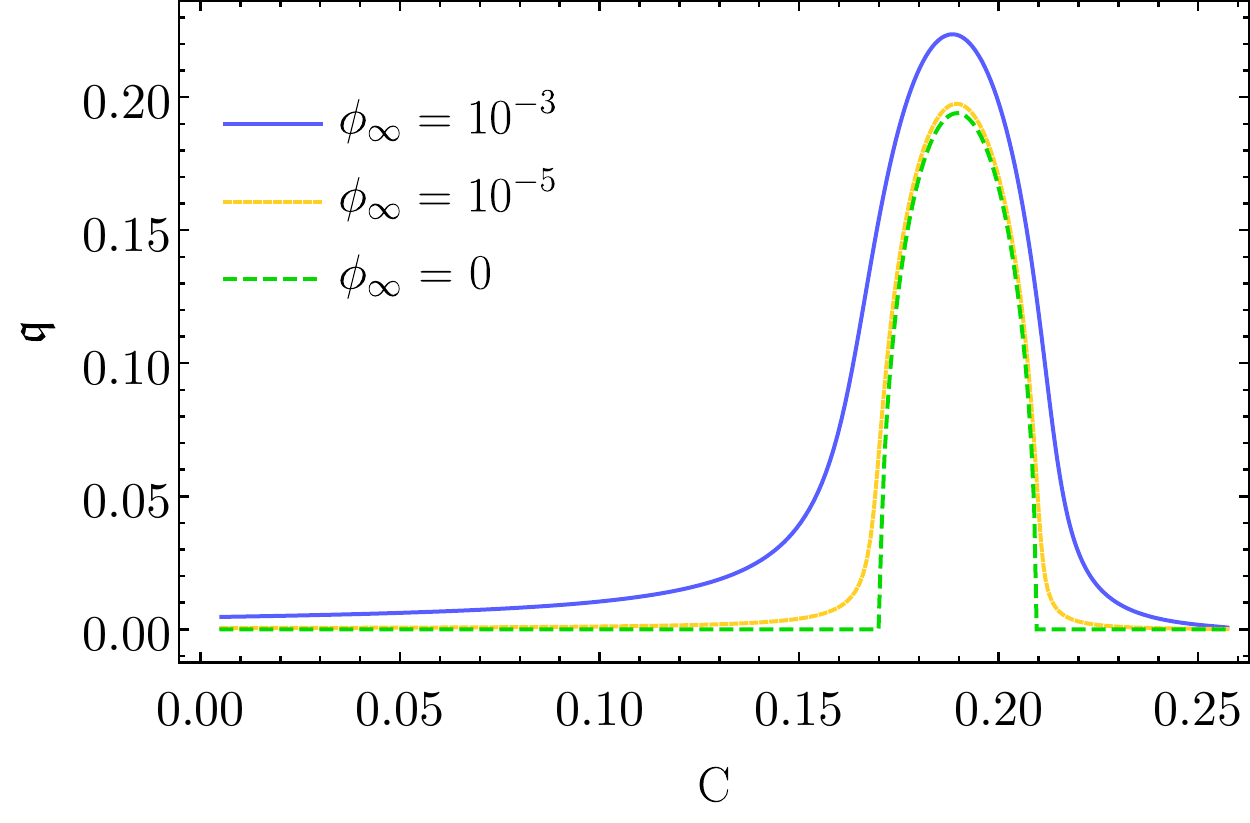}
  \caption{Scalar charge in units of ADM mass, $\mathfrak{q}$, versus compactness, $C$, for various values of the scalar field at infinity, for $\beta=-4.5$ and HB equation of state.}\label{fig:comp}
\end{figure}

\begin{figure}[t]
    \centering
    \subfloat[][Scalar charge in units of ADM mass, $\mathfrak{q}$ versus ADM mass for different values of $\beta$ with HB equation of state. \label{fig:alphabeta}]{\includegraphics[width=\columnwidth]{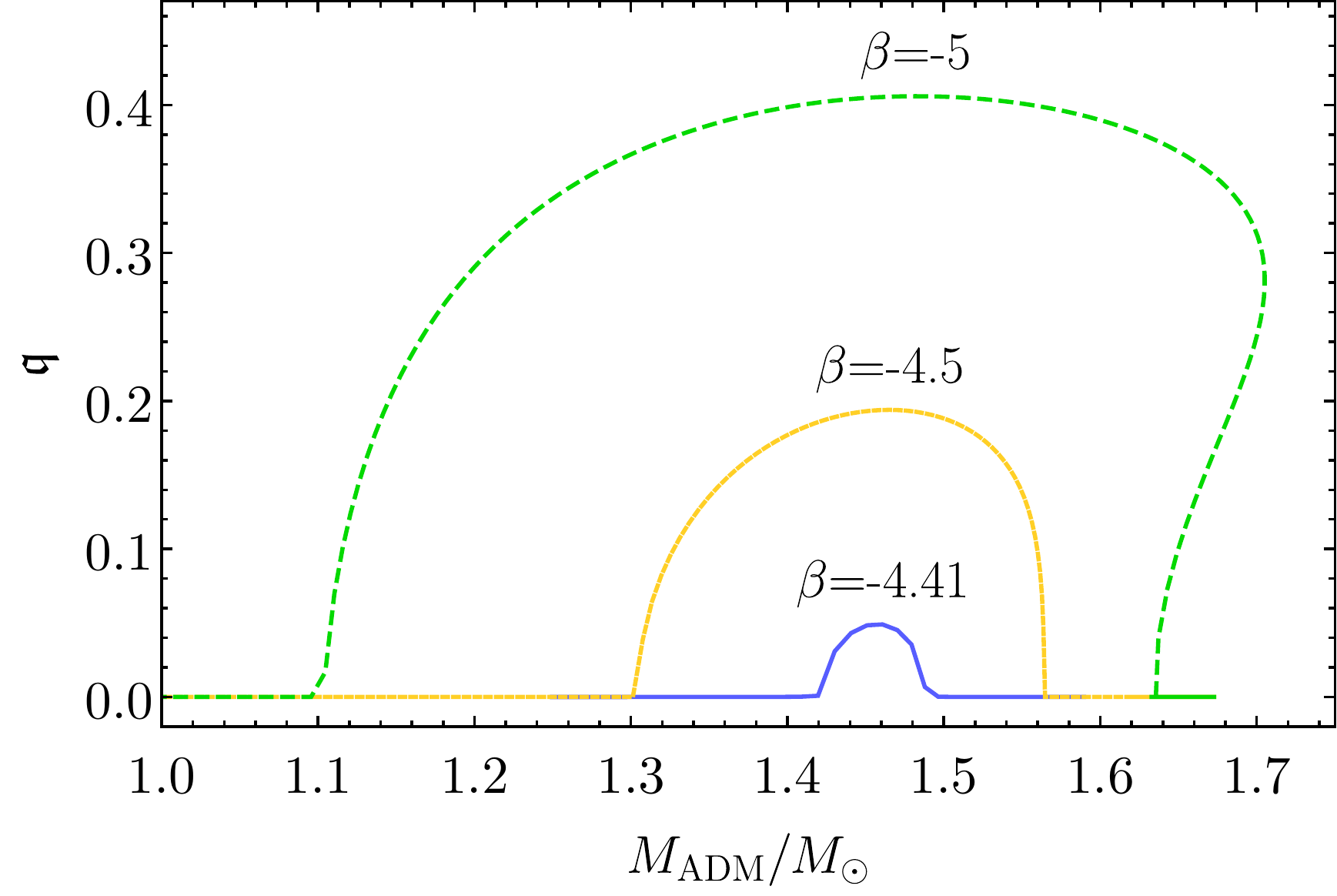}} \\
    \subfloat[][Scalar charge in units of ADM mass, $\mathfrak{q}$ versus ADM mass for different equations of state with $\beta=-4.5$. \label{fig:alphaeos}]{\includegraphics[width=\columnwidth]{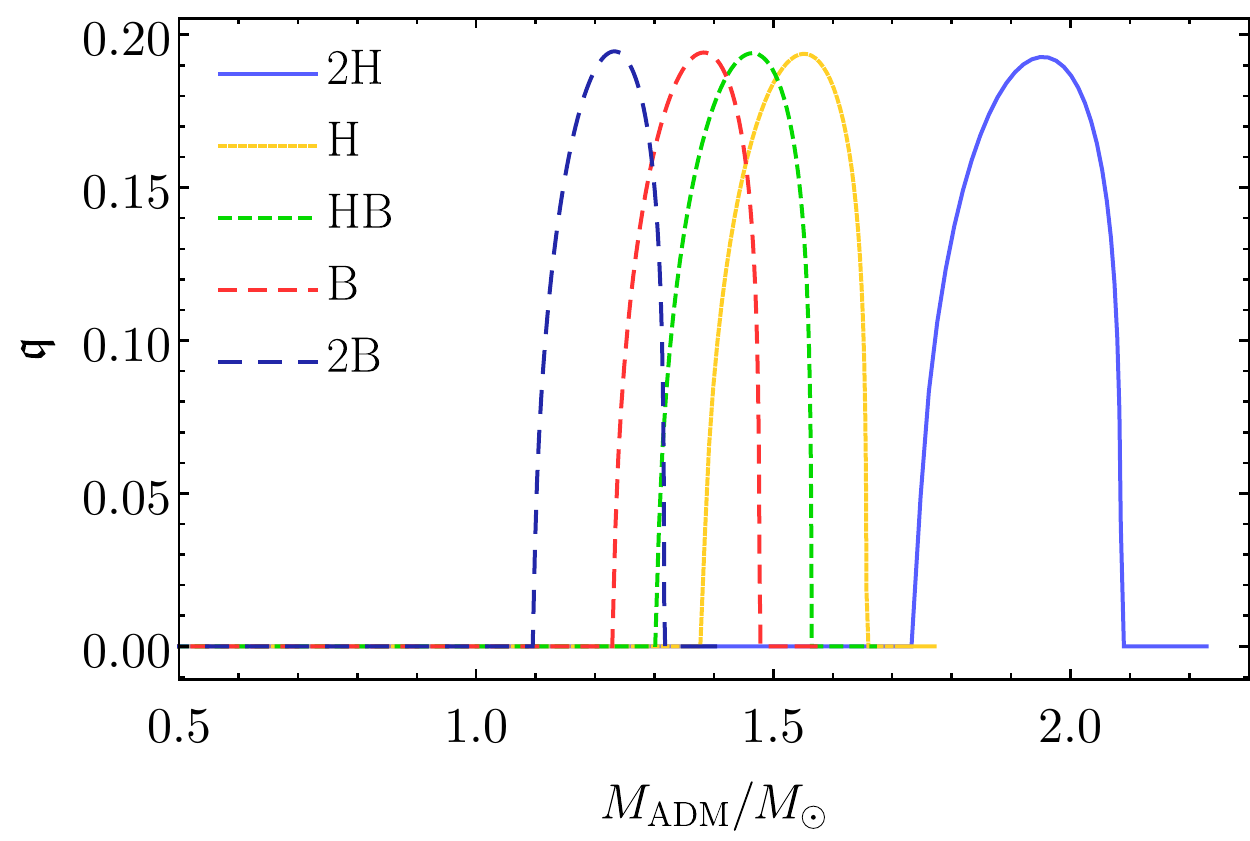}}
    \caption{Properties of scalarization for different solutions}
\end{figure}

In order to generate neutron star solutions we need to solve the system \eqref{eq:sistEM} numerically. It is a system of seven first-order equations, thus it needs seven initial conditions. Following Ref.~\cite{Damour:1993hw}, we chose $r=0$ as the starting point of integration, with initial conditions $\mu(0)=0$, $\nu(0)=0$, $p(0)=p_c$, $\phi(0)=\phi_c$, $\psi(0)=0$, $A_0(0)=0$ and $f(0)=0$. The last two conditions follow from the discussion in the previous section. In order to have a closed system, one needs to select an  equation of state (EOS) that relates the pressure $p$ with the  energy density $\epsilon$.

The integration is split into three different regions: the core, the crust and the exterior of the star. The core is the region where the number density satisfies  $n>n_c\,8.3\times10^{-5}$. For the core we have used one of five  polytropic EOS with different stiffnesses, namely 2H, H, HB, B and 2B as defined in Ref.~\cite{Read:2009yp}, with 2H being the stiffest and  2B being the softest.  The crust is the next region, where the pressure drops to zero. The condition $p=0$ defines the neutron star radius $r_s$. The EOS in the crust is chosen to be the same for all solutions as defined in Ref.~\cite{Read:2009yp}. In the exterior of the star, $p=\epsilon=0$. We consider the piece-wise polytropic modelling to be sufficiently realistic for our purposes.

\begin{figure*}
    \centering
    \subfloat[][Photon mass profile inside the neutron star and for large radii. \label{fig:mphot}]{\includegraphics[width=\columnwidth]{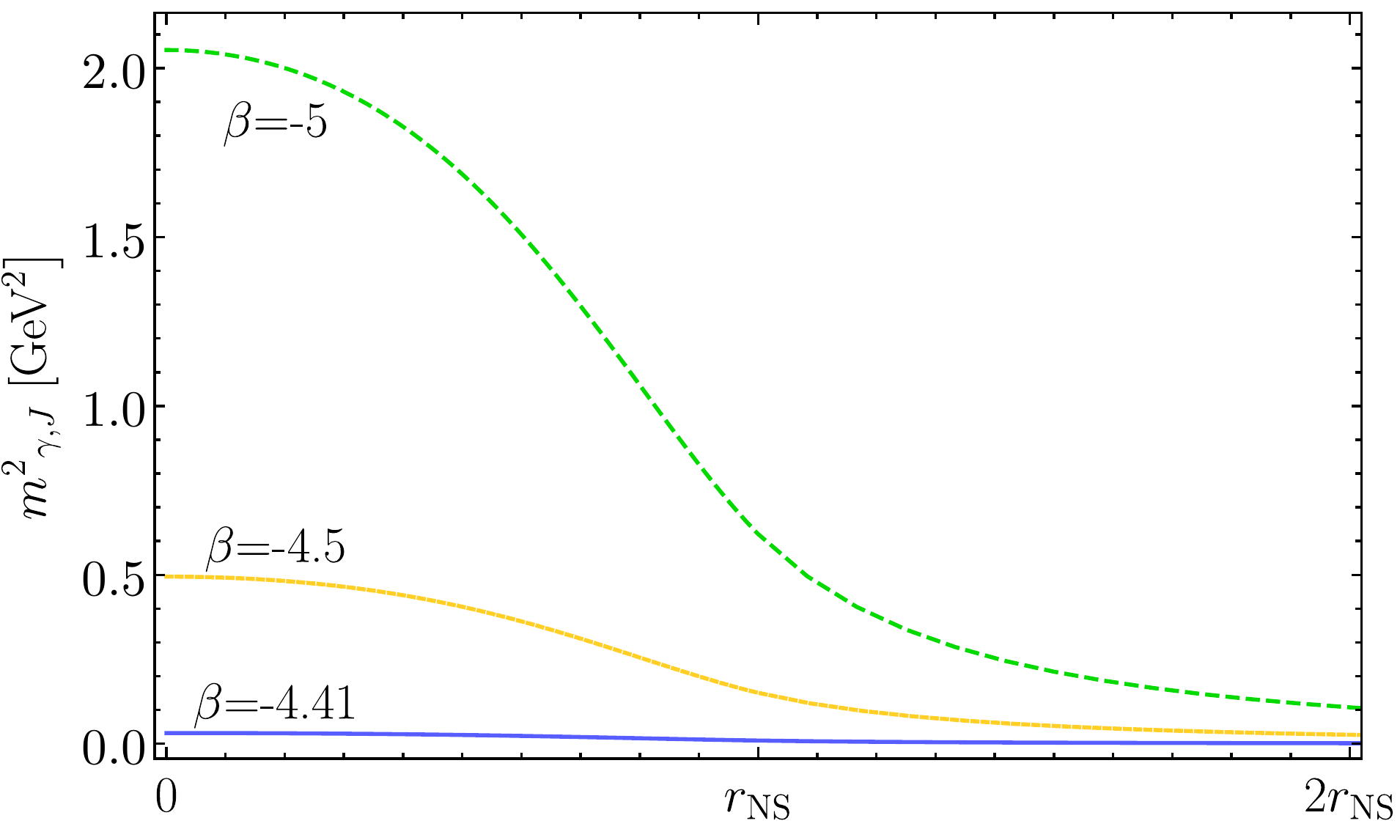}}
    \quad
    \subfloat[][Logarithmic profile of the photon mass in the Jordan frame for large radii. The style of the curves is the same as in the left panel. \label{fig:logmphot}]{\includegraphics[width=\columnwidth]{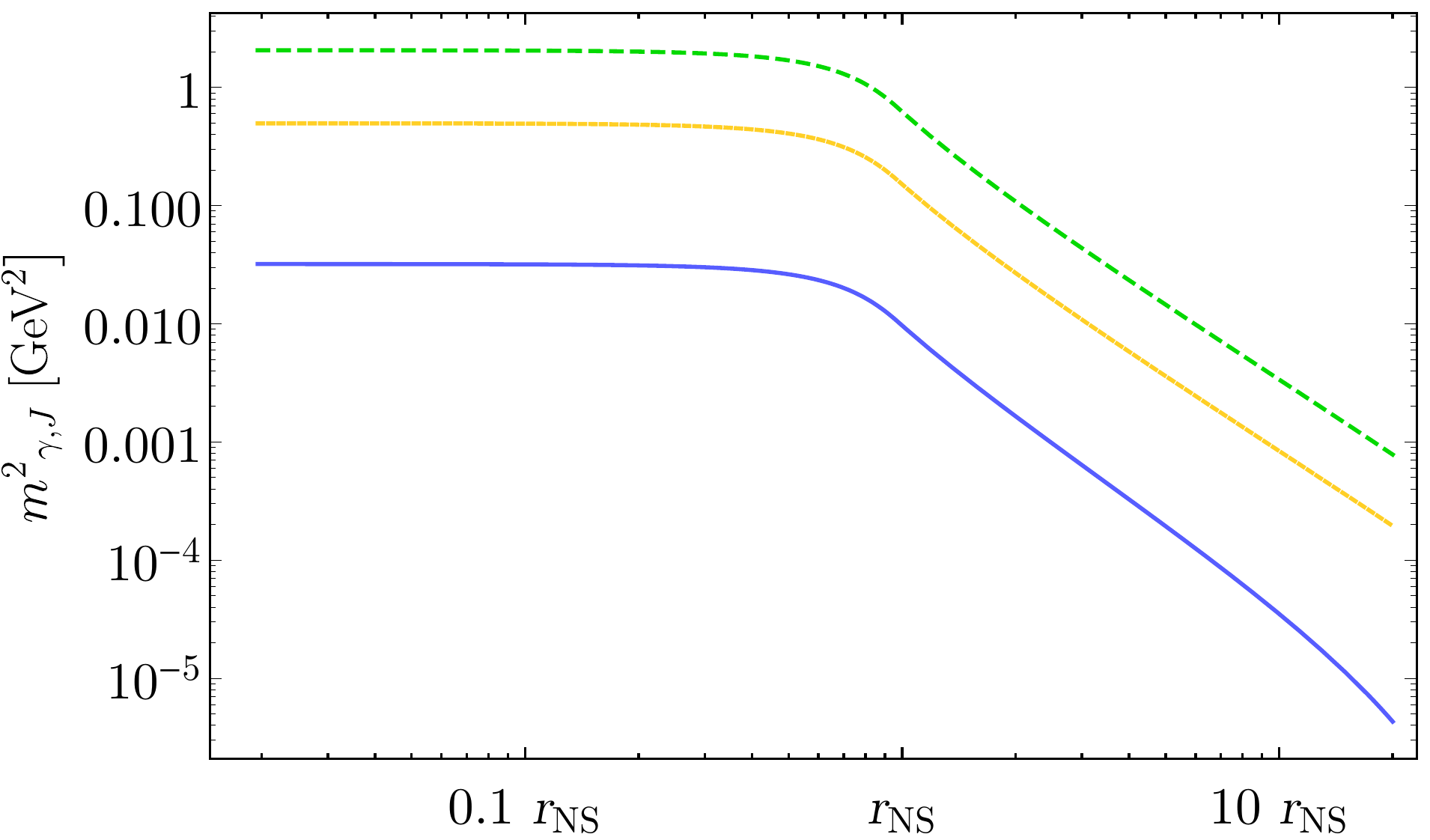}} \\
    \caption{Radial profile of the photon mass in the Jordan frame for different values of $\beta$, the HB equation of state, and for the coupling charge  $e=10^{-36}\,\text{C}$. The radial coordinate is normalized to the star radius.}
\end{figure*}

One can read-off the the ADM mass $M_\note{ADM}$ and the asymptotic charge for the scalar field $\mathcal{Q}$ through  asymptotic expansions of $g_{tt}(r)$ and $\phi(r)$:
\begin{align}
g_{tt}(r)=-1+\frac{M_\note{ADM}}{2\pi r}+\mathcal{O}\left(\frac{1}{r^2}\right) \label{eq:ADMmass} \\
\phi(r)=\phi_\infty+\frac{\mathcal{Q}}{4\pi r}+\mathcal{O}\left(\frac{1}{r^2}\right) \label{eq:scalarcharge},
\end{align}
where $\phi_\infty$ is the asymptotic value of the scalar field. For our following analysis it is useful to define the dimensionless parameter $\mathfrak{q}\equiv\mathcal{Q}/M_\note{ADM}$. $M_\note{ADM}$ and $\mathfrak{q}$ can also be expressed in terms of quantities that are evaluated at the surface of the star \cite{Damour:1993hw,Damour:1992we} which are denoted with a subscript $s$:

\begin{gather}\label{eq:asymptotics}
\mathfrak{q}=\frac{2\psi_s}{\nu'_s}, \\
M_\note{ADM}=2\pi r_s^2\nu'_s\left(1-\frac{2\mu_s}{r_s}\right)^{1/2} e^{\cal A},\\
{\cal A}\equiv \left[-\frac{\nu_s'}{({\nu_s'}^2+4\psi_s^2)^{1/2}} \tanh^{-1}\left(\frac{({\nu_s'}^2+4\psi_s^2)^{1/2}}{\nu_s'+2/r_s}\right)\right]
\end{gather}

Our intention has been to generate solutions that span all the allowed values for $M_\note{ADM}$ but have very small values of $\phi_\infty$. This latter choice is dictated by the need to be consistent with Solar systems and other  constraints \cite{Damour:1996ke,Damour:1998jk,EspositoFarese:2004cc,Freire:2012mg}. The value of $\phi_\infty$ is related to the choice of $\phi_c$. Hence, in practice, we have generated solutions in the following way. For a given value of $p_c$, we have used a shooting method in order to select the value of $\phi_c$ that yields the desired value of $\phi_\infty$ within our numerical accuracy.\footnote{We have used \textsc{Mathematica} for our numerical integration and the function \textsc{FindRoot} to find the value of $\phi_c$ that gives the desired $\phi_\infty$.} We have repeated this process for several values of $p_c$ in order to generate solutions with different values of $M_\note{ADM}$.

In Fig.~\ref{fig:comp}, we show the parameter $\mathfrak{q}$ as a function of the compactness $C=M_\note{ADM}/r_s$, for three different values of $\phi_\infty$. In this case, we have fixed $\beta=-4.5$ and the EOS to be HB. One can clearly see that the scalarization process acts as phase transition after a certain compactness is reached. The transition is very sharp if we impose that the scalar field vanishes at infinity; a non-zero value for $\phi_\infty$ gives a smoother profile. It is worth emphasising that compactness is intimately related to  curvature, and hence one can say that the curvature determines the threshold for the phase transition.

Since the scalarization occurs for every choice of $\phi_\infty$, for the rest of our analysis we have taken $\phi_\infty=0$, as this leads to vanishing mass for the photon asymptotically.
In Figs.~\ref{fig:alphabeta} and \ref{fig:alphaeos} we show how spontaneous scalarization acts for different choices of $\beta$ (first panel) and different EOS (second panel). Both plots show the scalarization parameter $\mathfrak{q}$ as a function of $M_\note{ADM}$. We clearly see that a higher value  of $|\beta|$ gives rise to a stronger scalarization. The choice of  EOS appears to have no effect on the scalar charge but it does  determines  the maximum allowed mass of the star.

These above results are in full agreement with previous simulations, {\textit e.g.}~Ref.~\cite{Damour:1993hw}. We can now use these solutions to calculate the mass profile for the photon. In Figs.~\ref{fig:mphot} and \ref{fig:logmphot} we show the mass of the photon in the Jordan frame, $m_{\gamma,\note{J}}^2$. For the coupling charge we have chosen $e=10^{-36}\,\text{C}$, which is 17 orders of magnitude below the charge of the electron. This is sufficient to yield photon masses that reach the GeV range in the centre of the star. We stress that $m_\gamma$ scales linearly with $e$.

\section{Discussion}
\label{sec:disc}

We have studied spontaneous scalarization as a Higgs-like mechanism at the nonperturbative level. We focused on the model of Ref.~\cite{Coates:2016ktu} in which the scalar is complex and coupled to the electromagnetic field. We first showed that static, spherically symmetric stars will not be endowed with an electromagnetic field as long as they are composed of electrically neutral matter. That is, scalarization will not induce an electric charge, despite the coupling between the scalar and the electromagnetic field.

We have generated numerical solutions that describe stars composed of a perfect fluid. Spontaneous scalarization is unaffected by the coupling between the scalar and the electromagnetic field; hence, our results are in perfect agreement with the existing literature. However, in our model scalarization generates a mass for the photon. We have used our solutions to determine this mass or, more precisely, its radial profile.

Remarkably, the mass of the photon can be large in the interior and in the vicinity of the neutron star even for very small coupling charge between the scalar and the electromagnetic field and for values of $\beta$ that support very little scalarization. For example, a coupling charge of $10^{-36}$ C --- 17 orders of magnitude smaller than the charge of the electron --- led to a mass in the GeV range and the mass scales linearly with the coupling charge. Scalarization appears to be a particularly efficient mechanism to generate a photon mass. The small values of the  coupling charge required suggest that it could be possible to have a large enough mass to affect the microphysics of the star without producing observable deviations from the standard model in earth-based accelerators.

On the other hand, the mass also scales linearly with $\phi$ and hence it has a $1/r$ fall off asymptotically. This means that photons can  continue to be sufficiently massive to leave a detectable imprint at large distances from the star. As a rather extravagant example, choosing the coupling charge to have the value of the electron charge and assuming that the neutron star is as far as possible from the Earth within the observable universe would still produce a mass of the photon on Earth that would be  orders of magnitude above current experimental bounds. As discussed above, the coupling charge can clearly be many orders of magnitude smaller and still lead to significant effects within the star.

It would be particularly interesting to add a (bare) mass to the scalar field for two distinct reasons. First,  having a mass term would change the asymptotic behaviour and lead to a very rapid fall off for the scalar, and consequently for the photon mass. Second, it has been recently shown that adding a mass helps evade binary pulsar constraints \cite{Ramazanoglu:2016kul}.  Indeed, for a massless case the most stringent  bound on the coupling parameter is $\beta\gtrsim -4.35$ \cite{Antoniadis:2013pzd}. Moreover, as one approaches the minimum allowed value of $\beta$, numeral simulations suggest that scalarization switches off \cite{Harada:1997mr,Chiba:1997ms}). For the equations of state we used and with the accuracy limitations of our code we could not confidently claim that scalarization occurred for $\beta \gtrsim -4.41$. Adding a mass term can render values of $\beta$ much larger than the one we have considered compatible with current bounds.
Note that, for a given value of $\beta$,  the mass of the scalar cannot exceed a certain value, else it would actually suppress scalarization entirely (this also implies that one would encounter the usual naturalness problem for massive scalar fields in quantum field theory.)

Another interesting case is when the star is rotating: highly spinning neutron stars can develop spontaneous scalarization event for values of $\beta$ larger than $-4.35$ \cite{Doneva:2013qva}. However in this case, our proof of vanishing electromagnetic potential does not hold since the ansatz \eqref{eq:ansatzA} is not the most generic one in an axisymmetric setup. The generalization of the proof goes beyond the scopes of this paper, though we do expect that following the steps of Sec. \ref{sec:emfield} would give a similar result.

The recent gravitational wave observation of a binary neutron star merger \cite{PhysRevLett.119.161101} has put stringent constraints on the speed of gravitational perturbations, assuming no delays on the electromagnetic emission (see e.g. \cite{Baker:2017hug}). Interestingly, these constraints become more loose if we allow a nonzero mass for the photon, since in order to measure with accuracy the delay between gravitational and electromagnetic signal, one should know the photon mass profile around the star.

Before closing we would like to stress that  generating a mass for the photon via spontaneous symmetry breaking within  the standard model might require a more elaborate model than the one considered here in order to preserve consistency (see Ref.~\cite{Ruegg:2003ps} for a discussion). As discussed in Ref.~\cite{Coates:2016ktu}, the model we considered here should be viewed as a toy model whose purpose is to demonstrate the effectiveness of spontaneous scalarization as a Higgs-like mechanism in a simplified setting. Indeed, our results provide strong motivation for exploring more elaborate models, in which the scalar that undergoes scalarization couples with other standard model fields. Our findings also bear an intriguing interpretation: it is possible that scalar fields, or more generally new degrees of freedoms nonminimally coupled to the metric, could lead to deviations in the standard model in the strong gravity regime, and still remain undetected in the weak field regime. If this turned out to be the case, one potential consequence would be drastic departures from what we currently consider realistic equation of states for neutron stars.

\begin{acknowledgments}
The research leading to these results has received funding  from the European Research Council under the European Union Seventh Framework Programme (FP7/2007-2013) / ERC Grant Agreement n.~306425 ``Challenging General Relativity''.  T. P. S. would also like to acknowledge 527 networking support by the COST Action GWverse 528 CA16104.
\end{acknowledgments}

\bibliography{bibnote}

\end{document}